# Observation of Zone-Folded Acoustic Phonons in Terahertz Quantum Cascade Lasers using Picosecond Ultrasonics

Axel Bruchhausen[1*†], Mike Hettich[1], James Lloyd-Hughes[2], Milan Fischer[3], Mattias Beck[3], Giacomo Scalari[3], Jérôme Faist[3], and Thomas Dekorsy[1]

[1] *Department of Physics and Center for Applied Photonics, Universität Konstanz, Germany.*
[2] *Department of Physics, University of Oxford, United Kingdom.*
[3] *Institute for Quantum Electronics, ETH Zürich, Switzerland.*

We have investigated the time-resolved vibrational properties of terahertz quantum cascade lasers by means of ultra-fast laser spectroscopy. By the observation of the acoustic folded branches, and by analyzing the involved phonon modes it is possible to extract accurate structural information of these devices, which are essential for their design and performance.

The THz frequency range of the electromagnetic spectrum from 1 to 10 THz (300 to 30 µm) has not been exploited much. Certainly not because of the lack of interest, because the potential applications in fundamental research and technological fields are numerous, but mainly because of the lack of cheap, compact and practical sources of coherent terahertz radiation. The absence of this type of electromagnetic sources has been nicknamed as "the terahertz gap" [1,2]. Terahertz Quantum Cascade Lasers (THz-QCLs) have settled as one of the promising light emitting devices for this spectral region, and they might be the door to the enormous number of possibilities that the conception of such sources would provide.

Conventional Quantum Cascade Lasers (QCLs), working in the mid-IR, are commercially available. However, the scaling of the devices to the THz range turned out *not* to be an easy task. Many approaches and improvements of these devices have been tried since the first experimental demonstration [3], but due to additional challenges associated to the charge transport such as the dephasing of the involved excitations and mainly scattering with phonons, impurities and interface roughness, the proper work of THz-QCLs has at present only been possible for working temperatures below ~150K [4-9]. It is thus essential to understand and have a good structural characterization and thorough understanding of the involved phonon states to plan further improvements to the devices' design. In the present work we have studied and characterized the acoustic vibrational states in this kind of QCL structures by means of ultra-fast optical spectroscopy.

QCLs are basically unipolar devices and are based on inter-sub-band transitions in semiconductor multiple quantum wells. They consist essentially of a 3 level system as sketched in Fig.1(a). In such a device, the carriers are injected into an initial state (normally denoted as state 3). The actual radiative transition ($h\nu$) occurs from state 3 to state 2, and is followed by a non-radiative transition to the lowest state 1. The interesting point of the QCL-devices is that this latter state is used as a "carrier injector" for a second, third, etc. equivalent stages, forming a kind of "cascade", which gives the name to this device. The key issue for this kind of solid state laser is to achieve a fast and high population of the initial state 3 to achieve the population inversion, and a fast depopulation of the final radiative state 2. Within this, the major challenge is to obtain a large population inversion for a sub-band-system very closely spaced in energy (THz range), but with a still strong oscillator strength for the optical transition in order to obtain a significant gain [4]. Since the energy levels are very close, different factors (temperature, electron-phonon scattering, impurity-scattering, interface roughness, etc.) are determinant in the smearing out of the transitions, contributing to the loss of coherence of the carrier states, and acting against the proper operation of the device [4,6].

As mentioned before, there have been many approaches to achieve such a device experimentally [4,7]. The one studied here, and that probably shows more potential, is the one that combines a direct vertical optical inter-sub-band transition by the bound-to-continuum approach of the "active region" [8], together with a so-called *resonant phonon design* to depopulate the final state 2 of the optical transition to "drain" carriers to the ground state and at the same time serve as the injection module for the next step [9], and the *resonant coupling* of the different regions [4,7]. Figure 1(a) schematically shows the different steps, starting from the initial state 3, the radiative transition ($h\nu$), the resonant coupling to a degenerated state 2a, and the consequent non-radiative transition by emission of the optical phonon ($\hbar\omega_{LO}$) to the state 1, which is used as the injector state of the next stage by being resonantly coupled to the state 3'.

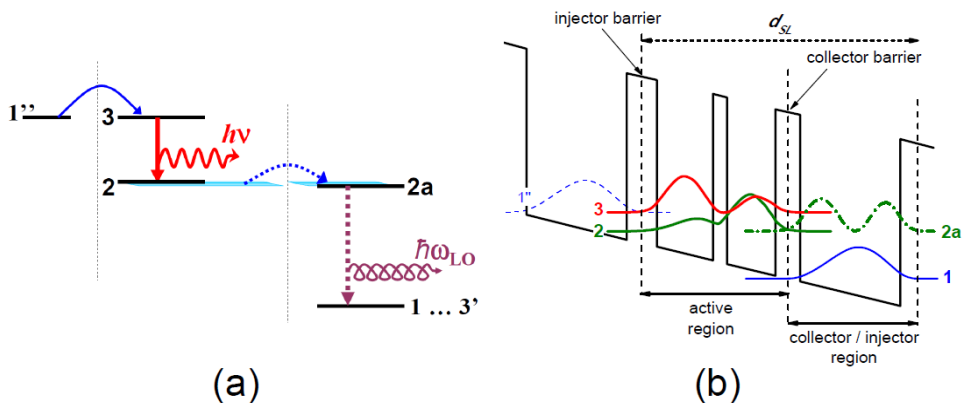

FIG.1: (a) Schematic representation of the electronic states involved in a THz-QCL. The radiative transition from state 3 to state 2 ($h\nu$), and the longitudinal optical phonon assisted transition from state 2a to state 1 ($\hbar\omega_{LO}$). (b) Schematic representation including the electronic conduction band structure along the growth direction and the squared electronic wave functions of the involved states of the multiple quantum wells. The different regions and the barriers are indicated. (Figures adapted from Refs.[4] and [5], respectively).

A more realistic sketch of Fig.1(a), including the quantum well picture of the conduction band, is shown in Fig.1(b). In this representation is clearly seen how essential the careful electron-state engineering needs to be performed is in order to accomplish the different steps. This careful tuning is mainly achieved by controlling the QWs' width, something that exposes the sensitivity of the design and the growth inaccuracies [4,5,7].

The QCLs have been grown by molecular beam epitaxy (MBE). We have analyzed a variety of different samples that show slight variations in the QCLs structure, all based on

GaAs/AlGaAs materials, but only the results corresponding to two of these samples will be shown in this work. A schematic illustration is shown in Fig.2(a) for sample #1 and in Fig.2(b) for sample #2. Starting from the substrate follows a parabolic grading, then the actual QCL active structure that basically is a superlattice (SL), and is finally cupped by an inverse parabolic grading.

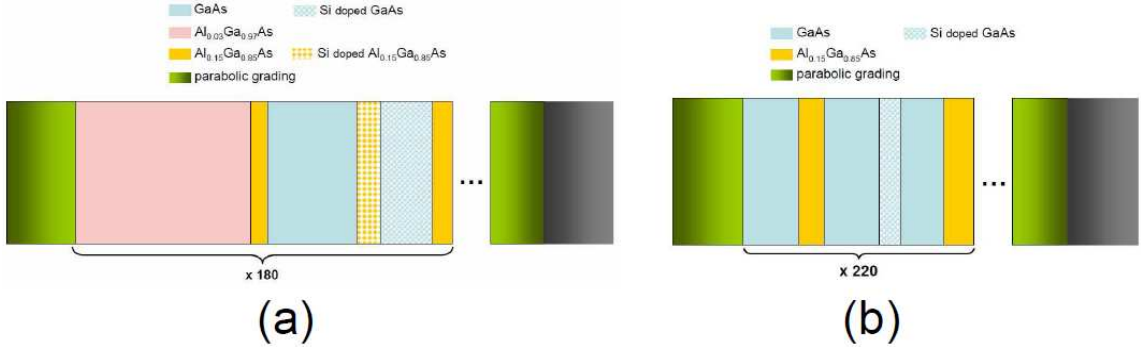

FIG.2: Schematic of samples studied: (a) sample #1 and (b) sample #2. Both samples are formed by a SL embedded within two parabolic gratings. For sample #1, the SL period is repeated 180 times and consists of the following layers: *32.5* / **3.2** / 16.6 / **4.3** / 10.0 / **3.7**, where the numbers are in given in *nm*, $Al_{0.03}Ga_{0.97}As$ layers are in italic, $Al_{0.15}Ga_{0.85}As$ layers in bold, the GaAs in roman; and the respective underlined layers correspond to $Al_{0.15}Ga_{0.85}As$ and GaAs doped with Si ($N_d \sim 2.05 \times 10^{16} cm^{-3}$ and $N_d \sim 2.4 \times 10^{16} cm^{-3}$ respectively). For sample #2, the SL period is repeated 220 times and consists of the following layers: 8.0 / **3.7** / 7.9 / **3.0** / 6.3 / **4.3**, where the underlined layer corresponds to GaAs doped with Si ($N_d \sim 1.0 \times 10^{17} cm^{-3}$).

These structures for working as QCL devices are processed (defining ridges) by lithography, wet etching, provided with (Cr/Au) contacts, and soldered on a (Cu) heat-sink [7]. Since our main interest is the actual core of the QCL structure, after the MBE growth the top parabolic grading was chemically removed, remaining the SL with a rather complex and relatively large period of $d_{SL} \sim 70.3 nm$ and $d_{SL} \sim 33.2 nm$ for samples #1 and #2, respectively.

Picosecond ultrasonics experiments have demonstrated to be an extremely suited technique to test the vibrational properties in a variety of structures [10-13], and since the vibrational states are extremely sensitive to slight structural changes it is possible by analyzing the vibrational properties to precisely derive structural parameters.

We have analyzed the samples by using the newly developed high-speed asynchronous optical sampling (ASOPS) pump-probe method [14,15]. This method basically uses two asynchronously coupled femtosecond Ti:sapphire ring lasers of $f_R \sim 800 MHz$ repetition rate. One of the lasers is used to provide the pump pulse, and the second to provide the probe pulse. The delay time between pump and probe pulses is accomplished by introducing a slight repetition-rate-offset of $\Delta f_R = 10 kHz$ between both lasers, which is kept constant through an active stabilization system [15]. As the time passes, this small offset in the repetition rate of both lasers provides the linear time delay of the probe pulse with respect to the pump pulse, which is used to resolve the temporal modulation of the probe-beams reflectivity ($\Delta R/R_0$) with 70*fs* time resolution. The absence of moving mechanical parts within the system allows for a very fast scanning rate together with a very high signal-to-noise ratio ($10^7$) in acquisition times of just a few seconds (see Ref.[15] for further details). All experiments have been performed in reflection geometry, focusing the beams down to a spot diameter of about $50\mu m$, and using typical pump (probe) average intensities of $200mW$ ($15mW$) and laser wavelengths centered at $790nm$ and at $810nm$ for pump and probe respectively. The time window of observation of $1.25ns$ is determined in our experiments by the time between two pump pulses, i.e. by the inverse of the laser repetition rate $f_R$.

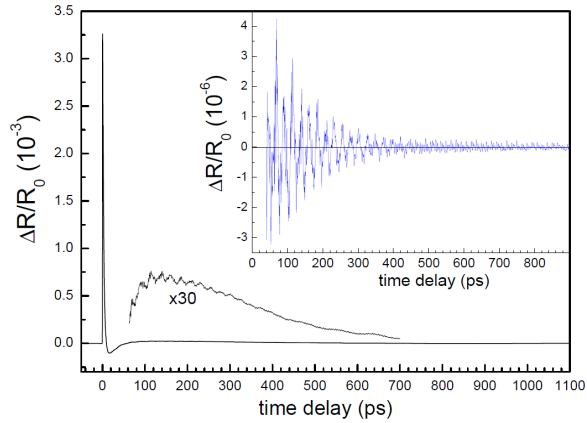

FIG.3: Typical time transient obtained for the probe beams reflectivity for sample #1. A portion of the curve is shown amplified by a factor of 30 in order to distinguish the reflectivity modulation due to the acoustic vibrations. The inset shows the oscillations after the subtraction of the electronic contribution. It can be observed that high frequency components are still present at long time delays.

A typical time transient of the temporal modulation of the reflected probe beam $\Delta R/R_0$ obtained by this technique is shown in Fig.3. The signal is characterized by a strong onset at $t=0ps$ due to the carrier excitation, followed by a multiple exponential-like decay. This signal is additionally modulated by small oscillations that correspond to the actual interest of this work, and which correspond to the modulation of the reflectivity due to the presence of the acoustic phonon modes, coherently generated in the structure by the pump pulse [10,14,15]. The inset of Fig.3 shows the oscillations after the subtraction of the electronic contribution. Note that the reflectivity scale has changed by an order of three.

By numerically Fourier transforming (FFT) the extracted signal, it is possible to identify the different active frequency modes that compose the temporal evolution. The spectral domain for the sample shown in Fig.3, can be observed in Fig.4 (bottom panel). Modes with frequencies up to ~1 THz can be clearly noticed. Figure 4 shows the analogous situation, but for sample #2.

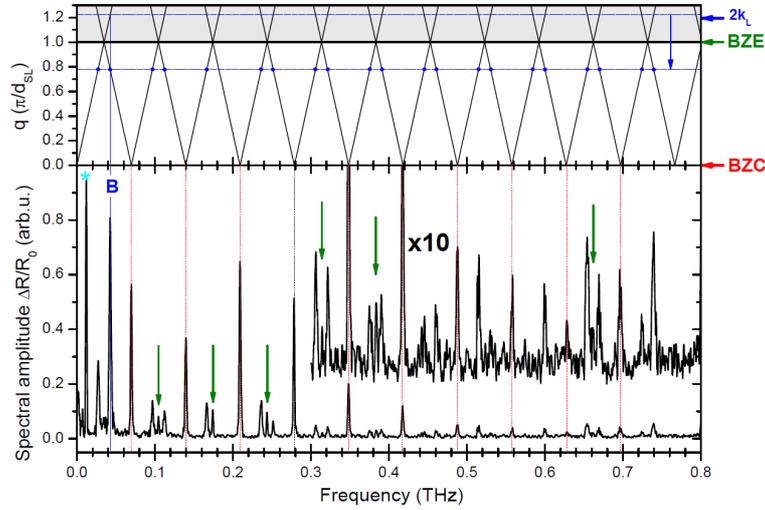

FIG.4: (top panel) calculated dispersion of the folded acoustic phonons for sample #1, according to the model described in the text. The layer widths have been reduced by a 3% with respect to their nominal value [see Fig.2(a)]. The gray zone corresponds to part of the second reduced Brillouin zone, and the Brillouin zone center (BZC) and Brillouin zone edge (BZE) are indicated. (bottom panel) Fourier transformation (FFT) of the measured time-resolved $\Delta R/R_0$ shown in the inset of Fig.3. The peaks marked with a "*" are an artifact. See text for further details.

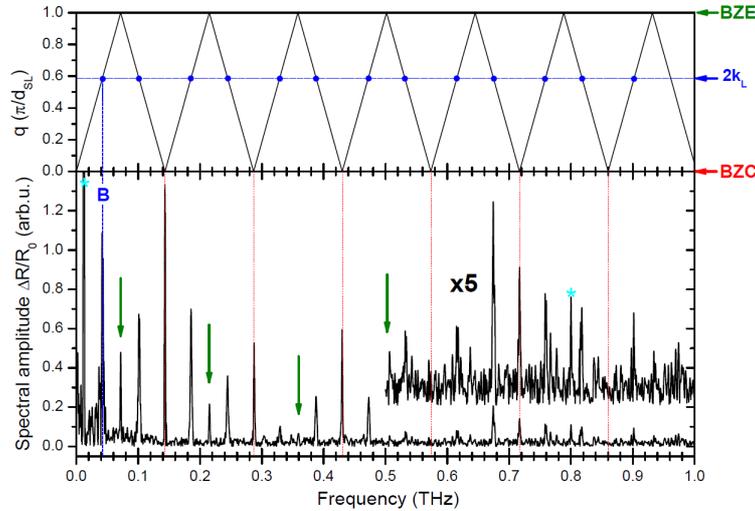

FIG.5: Analogous to Fig.4 but for sample #2. (top panel) dispersion of the FA-Phonons. The layer widths have been reduced by a 1% with respect to their nominal value [see Fig.2(b)]. (bottom panel) Fourier transformation (FFT) of the measured time-resolved $\Delta R/R_0$ for sample #2. The peaks marked with a "*" are artifacts. See text for further details.

In order to understand the origin of these sharp peaks and to comprehend the acoustic modes involved we have analyzed the system based on the simple model proposed by S. M. Rytov to describe acoustic modes in an infinite periodic system [16]. As mentioned above, the core of the QCL is basically a complex SL composed by 4 to 8 intrinsic and doped layers, depending strongly on the electronic states design and on the approach used to collect and extract the carriers from the active region. Due to the super-periodicity imposed to the acoustic phonons when propagating through the SL's structure, the acoustic branches of the material are folded back into a new effective reduced Brillouin zone (BZ) [17]. At the reduced Brillouin zone center (BZC) for $q=0$ and at the reduced Brillouin zone edge (BZE) for $q=\pi/d_{SL}$, as a consequence of the modulation caused by the acoustic impedances, i.e. the different relations between the material's sound velocities ($v_i$) and the thicknesses ($d_i$) of the layers, the dispersion degeneracy is lifted and so called "mini-gaps" are opened [17-20]. It turns out that for the kind of samples analyzed here, partially as a consequence of the complex layered period, the width of the mini-gaps result much smaller than the actual resolution of our system determined by the laser's repetition rate. The reduced acoustic frequency dispersion obtained by the extension of the Rytov model for SLs containing many layers per period, neglecting the acoustic impedance modulation, yields

$$f_m(q) = \pm \frac{1}{2\pi} q v_{eff} + m \frac{v_{eff}}{d_{SL}}, \qquad (1)$$

where $v_{eff}$ is the effective acoustic sound velocity given by $v_{eff}^{-1} = \frac{1}{d_{SL}} \sum_i \frac{d_i}{v_i}$, $d_i$ and $v_i$ are the corresponding thickness and acoustic sound velocity of the layer $i$, $d_{SL}$ is the SL period, and $m$ is an integer number ($m=0,1,2...$) that determines the folded acoustic branch. The calculated folded dispersion given by Eq.(1) for the analyzed samples is shown in the top panels of Fig.4. To calculate the acoustic dispersions the well established bulk parameters for GaAs and $Al_xGa_{1-x}As$ have been used [21].

Contrary to what is the case in Raman spectroscopy [17], where the active modes depend on the scattering geometry, due to the generation and detection processes involved, the active modes (i.e. the experimentally observable modes) in pump-probe experiments are a combination of modes with effective wave-vectors of $q=0$ and those modes corresponding to a wave-vector $q=2k_L$, where $k_L$ is the wave-vector of the respective probe laser light [12,22-24]. This gives rise to characteristic "triplets" formed by a center peak with $q=0$ and two symmetric peaks towards higher and lower frequencies for $q=2k_L$. Since the composition of each layer is well known from the MBE growth method, the frequencies of the observed acoustic triplet can be fitted to the calculated branches by varying the individual layers width $d_i$ of the SLs. Starting from the nominal values for $d_i$ (indicated in Fig.2), the best agreement is obtained for a uniform reduction of the $d_i$ values of 3% and 1% for sample #1 and sample #2 respectively, in good agreement with the results provided by X-ray diffraction experiments. It can be observed comparing the top and bottom panels in Fig.4 and Fig.5 that the frequency of all appearing modes can be well assigned to the corresponding $q$ in the folded acoustic phonon dispersion. Those modes with $q=0$ are indicated with a vertical (red) dashed line, and the experimentally observed peaks ascribed to $q=2k_L$ are indicated with a (blue) dot in the corresponding top panel where the phonon dispersion intersects the horizontal (blue) line representing this effective phonon wave-vector (see indicated blue arrow). The peak indicated with "B" and the vertical dotted (blue) line corresponds to the so called Brillouin or bulk-like peak. Note that for sample #1 in Fig.4, since the SL's period is relatively large, the value of $2k_L$ exceeds the first BZE wave-vector ($q=\pi/d_{SL}$) by a 22%, and falls into the second reduced BZ (shadowed area in Fig.4, top panel). Therefore the effective $q=2k_L$ wave-vector is additionally folded into the first BZ. Due to this folding a further peak is observed below the actual Brillouin peaks frequency.

The higher acoustic branches are much more sensitive to structural variations [17, 25]. Thus, the fact that peaks corresponding to up to 22 folded acoustic branches for sample #1 and up to 14 for sample #2 are observed, is a strong indication of the good periodicity and interface quality of the samples [25]. Additionally by fitting the frequencies of higher order triplets greater accuracy is achieved for the derived structural parameters.

An interesting point to notice, is the systematic apparition of additional peaks (indicated by the vertical green arrows) that quite precisely match the BZE frequencies at $q=\pi/d_{SL}$. The reason of these BZE modes to become active is not clear yet, and further investigations need to be performed to elucidate this point. Nevertheless it is important to notice that the excitations (pump) as well as the detections (probe) energies are near and above the electronic states, and effects of an electronic resonant excitation maybe the cause of the symmetry breakdown and consequent observation of these peaks.

In summary, we have analyzed the core of THz-QCL structures, and using the high-speed asynchronous optical sampling pump-probe method we have investigated the time-resolved dynamics of the acoustic modes that exist within these devices. The analysis of the Fourier components of the probe-beams reflectivity modulation due to the presence of the acoustic vibrations and the comparison to a simple Rytov model permits an accurate determination of the structural parameters, in particular the precise layer thickness, resulting in deviations from the nominal values of -3% and -1%. Up to date the inter-sub-band LA phonon scattering rates are normally calculated assuming bulk like phonon dispersions [6]. The determination of the correct structure, allows a more truthful description of the acoustic states, which is essential for calculating these rates in a more rigorous way. Since the acoustic folded vibrations have a finite population even at low temperatures, they may influence the optical line width and the device transport by changing the in-plane scattering rate.

**ACKNOWLEDGEMENTS**

A. Bruchhausen thanks the Alexander von Humboldt Foundation (Bonn, Germany) for its financial support. This work is partially supported by the Deutsche Forschungsgemeinschaft (DFG) through the SFB 767 and by the Ministry of Science, Research and Arts of Baden-Württemberg, Germany.